\documentclass[utf8,12pt]{article}
\usepackage[utf8]{inputenc}
\usepackage[english]{babel}
\pdfoutput=1
\usepackage{graphicx}
\usepackage{xcolor}
\usepackage{caption}
\usepackage{subcaption}
\usepackage{authblk}
\graphicspath{{images/}{./images/}}
\usepackage{blindtext}
\usepackage{float}
\usepackage{subfiles} % Best loaded last in the preamble

\usepackage{url,lineno,microtype}
\usepackage[onehalfspacing]{setspace}

\title{3D Structure from 2D Microscopy images using Deep Learning}
\begin{document}

\author[1]{\small Benjamin J. Blundell}
\author[2]{\small Christian Sieben}
\author[3]{\small Suliana Manley}
\author[4]{\small Ed Rosten}
\author[1]{\small QueeLim Ch'ng}
\author[5]{\small Susan Cox}

\affil[1]{\footnotesize Centre for Developmental Biology, Institute of Psychiatry, Psychology \& Neuroscience, King's College London, SE1 1UL, UK}
\affil[2]{\footnotesize Nanoscale Infection Biology Lab (NIBI), Helmholtz Centre for Infection Research, Germany}
\affil[3]{\footnotesize \'Ecole Polytechnique F\'ed\'erale de Lausanne, Switzerland}
\affil[4]{\footnotesize Snap, Inc., 7--11 Lexington St, London W1F 9AF, UK}
\affil[5]{\footnotesize Randall Centre for Cell \& Molecular Biophysics, King's College London, SE1 1UL, UK }

\maketitle

\begin{abstract}
Understanding the structure of a protein complex is crucial in determining its function. However, retrieving accurate 3D structures from microscopy images is highly challenging, particularly as many imaging modalities are two-dimensional. Recent advances in Artificial Intelligence have been applied to this problem, primarily using voxel based approaches to analyse sets of electron microscopy images. Here we present a deep learning solution for reconstructing the protein complexes from a number of 2D single molecule localization microscopy images, with the solution being completely unconstrained. Our convolutional neural network coupled with a differentiable renderer predicts pose and derives a single structure. After training, the network is discarded, with the output of this method being a structural model which fits the data-set. We demonstrate the performance of our system on two protein complexes: CEP152 (which comprises part of the proximal toroid of the centriole) and centrioles.
\end{abstract}

\section{Introduction}

Imaging mesoscale 3D biological structures (that is, those between the nano- and the micro-scale) is a critical problem in biology, as many processes of biological interest rely on collections of proteins or other molecules arranged into a distinct architecture. Currently two major techniques can provide data on the shape of such aggregates: electron microscopy and light (particularly fluorescence) microscopy. Electron microscopy (EM) offers resolution below 1~nm, but is limited in the thickness of the samples it can observe, and analysis is relatively complex, generally requiring multiple particle averaging~\cite{milneCryoelectronMicroscopyPrimer2013}. Fluorescence microscopy is experimentally relatively simple and can deal with larger samples, but generally yields only single images which are limited in resolution to about 250~nm~\cite{schermellehGuideSuperresolutionFluorescence2010}.

Super-resolution techniques allow this limit to be broken, pushing the achievable resolution down to 20-100~nm. In particular, single molecule localisation microscopy (SMLM) yields high resolution images (around 20-30~nm), while allowing large amounts of data to be collected \cite{schermellehGuideSuperresolutionFluorescence2010,holdenHighThroughput3D2014} and being relatively experimentally simple.
SMLM imaging has a trade off between the $x,y$ and $z$ resolution: gaining information in the $z$ direction is possible, but generally at the expense of in-plane information quality~\cite{badieirostamiThreedimensionalLocalizationPrecision2010}. Therefore, 2D images will have the highest localisation quality, but clearly limit information on 3D structure.

The challenge of how to infer 3D information from 2D images has been tackled both from the perspective of synthesising EM images to create a 3D structural model~\cite{milneCryoelectronMicroscopyPrimer2013}, and in the computer vision field to infer a 3D structure from a single image of a single object~\cite{fanPointSetGeneration2017}. In recent years, deep learning has emerged as a promising approach to improve structural fitting.

Convolutional neural networks are one of the most well known forms of Deep Learning -  convolving the data with a kernel~\cite{goodfellowDeepLearning2016}. This process reduces the size of the principal data dimensions, creating a number of feature maps or filters, each sensitive to a particular, local aspect of the data. Through training, the network parameters adjust to produce the required output.

Here, we use a deep learning network to infer the pose of point cloud data and 3D structure. Our algorithm HOLLy (Hypothesised Object from Light Localisations) allows us to perform a completely unconstrained model fit from 2D SMLM images.

\section{Methods}
\subsection{Modelling pose using deep learning}
HOLLy fits a 3D model against a set of 2D images of the same biological structure. The input images are typically super-resolved SMLM reconstructions, each of which is a z projection of the structure being imaged from some unknown rotational orientation and translation. The goal is to deduce the pose (rotation and translation) for each input image and infer a single 3D model for the entire data-set.

The 3D model is a collection of points (with their co-ordinates represented by a matrix) which are initiated at random positions. The current positions of the points, and the pose corresponding to each input image, are used to generate a simulated microscopy image corresponding to each input image (with the image being projected in z into a single x-y plane). The image is rendered with a Gaussian at each point, as is standard for SMLM. Each Gaussian has the same sigma, which is a parameter of the renderer, and the resulting image is differentiable with respect to the point coordinates and sigma. Our renderer is designed to efficiently and accurately render SMLM point clouds. This is in contrast to existing state of the art such as OpenDR~\cite{loperOpenDRApproximateDifferentiable2014a}, DiRT~\cite{hendersonLearningSingleImage3D2020a}, PyTorch3D~\cite{ravi2020pytorch3d}, Pulsar~\cite{lassner2020pulsar} and DWDR~\cite{hanDRWRDifferentiableRenderer2020} which are designed primarily to render illuminated, textured meshes with perspective cameras (or in the case of Pulsar and \cite{insafutdinovUnsupervisedLearningShape2018}, rendering with spheres), our renderer is simpler and more closely models SMLM. Rather than rendering rasterised triangles, HOLLy converts the final 2D points to Gaussians. 

We used a simple convolutional neural network (CNN) consisting of 10 layers of strided convolutions and Leaky-ReLU \cite{aggarwalNeuralNetworksDeep2018}, followed by two fully connected layers. Figure~\ref{fig_network} highlights the major components (further information can  be found in Supplementary Data: HOLLy technical details).

The CNN yields six outputs. The position and orientation of the model are described by the translation in $X$ and $Y$, and 3 rotation parameters for which we used the axis-angle formulation. The sixth parameter is the output-sigma value which is used as the sigma for the renderer. Note that the output-sigma explicitly differs from the resolution of the input images, i.e.\ in the case where input images are themselves reconstructions of SMLM data, the sigma used for their reconstruction (\emph{input-sigma}) is not the same as output-sigma.

In principle, if the input data were perfect, the output sigma could be fixed to be equal to the input sigma. Since this is not the case, allowing the model to predict output-sigma allows it to account for some of the noise in the data. For example, consider the case of scatter (noise in the position of fluorophores). That is essentially a stochastic blur of the model structure, so when the reconstructed 3D model (which has no scatter) is rendered, the sigma needs to be higher in order for the output to be a good match to the input.
This is discussed further in Section \ref{sec:sigma}.

\begin{figure}
\centering
\includegraphics[width=0.8\textwidth]{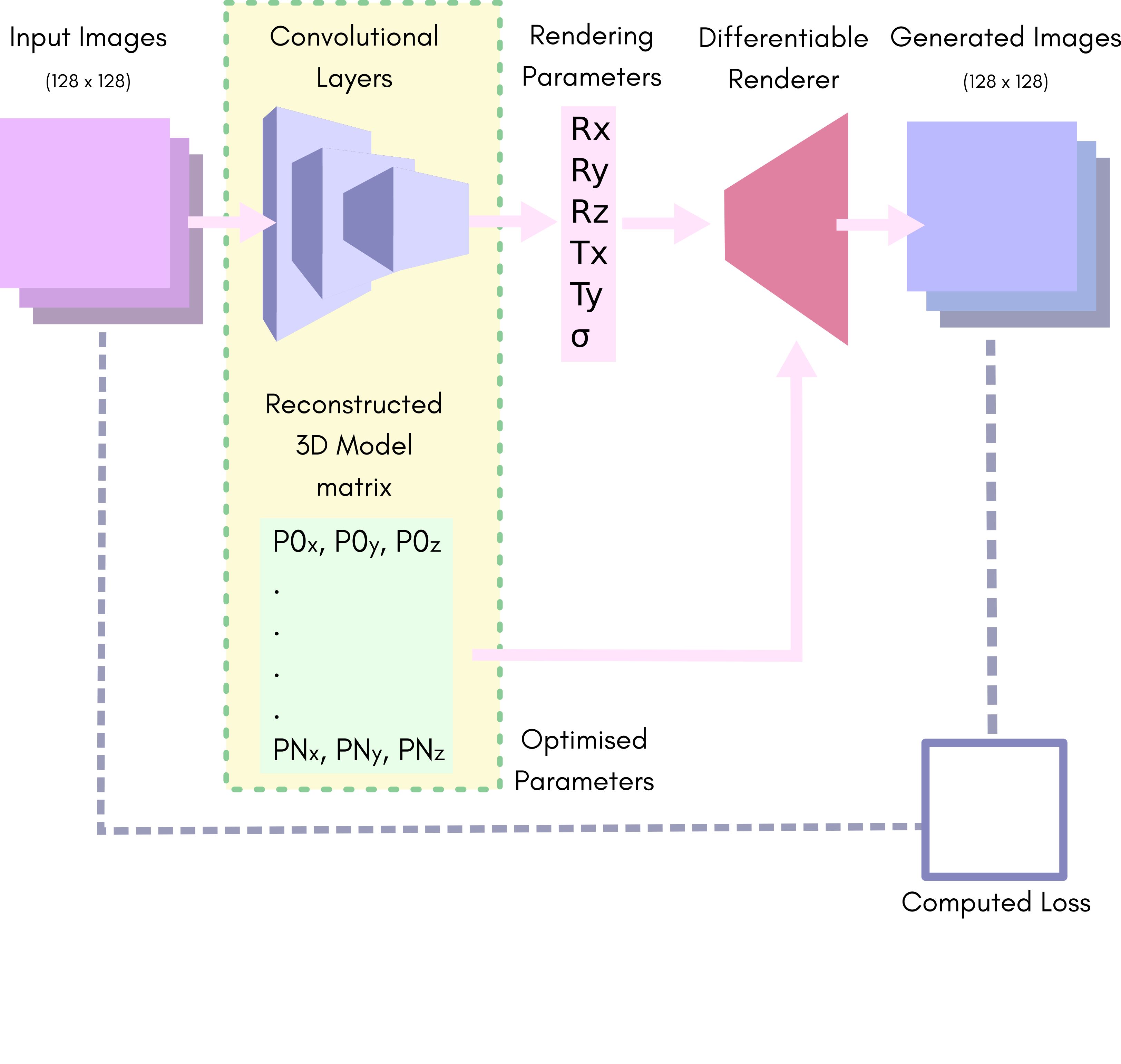}
\caption{An overview of our network. Batches of images of size 128 x 128 pixels are fed to the convolution layers, which reduce the batch down to 6 parameters. These are passed to the rendering pipeline along with the 3D reconstruction matrix to produce a batch of output images. The rendering parameters Rx, Ry and Rz represent the rotation in \lq Angle-Axis\rq form. Tx and Ty represent translation in the X, Y plane. S represents the predicted output-sigma. The 3D reconstruction matrix contains a list of vertices representing the predicted point-cloud.}
\label{fig_network}
\end{figure}

The key element of our system is the use of a CNN to predict the pose for each input image. Allowing for a pose per image is a significant advantage over techniques such as template matching based Cryo-EM \cite{milneCryoelectronMicroscopyPrimer2013}, or classification of the images by view \cite{salasAngularReconstitutionbased3D2017} since the system is not limited to a small number of orientations, and views do not have to be determined a-priori on unknown structure in order to build a classifier.

\begin{figure}
\centering
\includegraphics[width=0.5\textwidth]{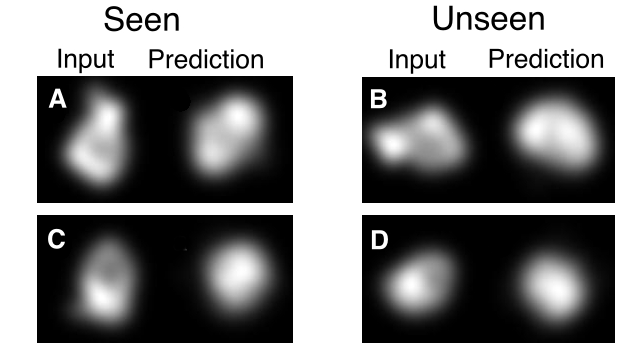}
\caption{Demonstration of information sharing between different poses with the CNN. HOLLy was stopped half-way through the first epoch of training, meaning only half of the data has been used for training.
Note that for the half of the data which has been seen, this corresponds to a single step of gradient descent per image, and half of the data has not yet been used at all.
Already it can be observed that in many cases on seen data (illustrated in A, C) there is correspondence between the input and output shapes (albeit imperfect as it is very early in the optimization process). The advantage of the CNN can be observed in the results on unseen data where this correspondence also exists (B, D). This partial convergence on unseen data shows that the CNN allows earlier data to assist in the convergence of data seen later, which provides a very substantial improvement over modelling poses separately.
\label{fig_smooth}}
\end{figure}

Additionally, using a CNN to predict the pose has a big advantage in modelling a pose per image as it makes the overall optimization much more tractable. The reason for this is that the space is in some sense smooth and images that are close in appearance will usually also be close in pose. This allows the network to aggregate information from similar images in order to get a better prediction of the pose for all of them. It also allows for fast convergence because an improvement on one image can cause an improvement in many others. We illustrate this in Figure~\ref{fig_smooth}, where data that is not seen during training can generate outputs that correspond to the input.

The advantage of using a CNN can be illustrated by attempting to solve the same problem by direct optimisation. We removed the convolutional layers from the architecture shown in Figure~\ref{fig_network}, replacing them with a single $5\times N$ matrix ($N$ being the size of the training set). A training batch consists of a batch of images and their corresponding poses from that $5\times N$ matrix. These differentiable render is used to render the model with these poses.
This rendered images are compared to the corresponding input image creating a loss as before. The loss is back-propagated through the diffentiable renderer and used to update the model and the poses. Various learning rates, models and optimisers were tested.

This direct optimisation approach could not reproduce 3D sample structure or model the pose correctly. We suspect this is due to both a lack of shared rotational model between data and the difficulty of modelling rotation. The results can be found in Supplementary Data: Direct Optimisation. These results demonstrate the advantages of using a CNN in this scenario.

\subsection{The output is a structural model rather than a trained network}
Often, the value of a neural network is the network itself that can be used to predict, discriminate or otherwise solve a particular problem once trained. Our approach ignores the network once it has been trained; the value in our approach is the 3D model stored in the Reconstructed 3D Model matrix.

This 3D model gradually improves as training continues. The user can stop training at any time, typically when the loss stops improving. The final positions of the points in the 3D reconstruction matrix represent the final structure, whereupon the network is no longer required.

\subsection{Simulated Data Models}

In order to evaluate HOLLy, we selected a number of ground-truth point-clouds with different characteristics: a reduced version of the Stanford Bunny\footnote{\label{fn_bunny}\url{http://graphics.stanford.edu/data/3Dscanrep/\#bunny}}, the Utah Teapot\footnote{\label{fn_utah}\url{https://www.computerhistory.org/collections/catalog/102710359}} and an approximation of the CEP152/HsSAS-6 complex~\cite{siebenMulticolorSingleparticleReconstruction2018}. 

All of the models consist of a relatively small number of vertices (fewer than 400). Each have unique characteristics, such as different numbers of vertices, symmetries and voids (see Section \ref{sec:results} - Results). The Stanford Bunny and Utah Teapot are standard in computer vision tests as they have properties that are likely to prove challenging. The Utah teapot is close to, but not quite, symmetric, and the Stanford bunny has fine structure (ears) but also relatively large areas of smooth structure (back). These properties showcase the potential of the method to yield results as experimental data improves.

As we have complete control of data-set generation from synthetic models, we must choose the distribution of data across the translation and rotation space. We uniformly sampled the 3D rotation group---SO(3)---which consists of all rotations in Euclidean $\Re^3$ space, centred at the origin, using the equation presented by \cite{kirkGraphicsGems1994}\footnote{ \label{fn_rotate}\url{https://demonstrations.wolfram.com/SamplingAUniformlyRandomRotation/}}. The models used in the simulated data are small enough to be rendered \lq on-the-fly\rq\ into images as the network trains.

\subsection{Experimental data from biological structures}

Our main biological targets were a centriolar complex comprised of the CEP152 protein, and purified centrioles.

The first dataset was a super-resolution (STORM) microscopy data-set of CEP152, obtained and analysed as described in \cite{siebenMulticolorSingleparticleReconstruction2018}. The structure of this centro-symmetric complex has been fitted with a toroid and found to be 400~nm in diameter \cite{siebenMulticolorSingleparticleReconstruction2018}, which subsequent work confirmed \cite{kimMolecularArchitectureCylindrical2019}. This  yielded a list of localisations for each identified CEP152 structure which were reconstructed localisations into 2D images, rendering with a Gaussian. Since the number of localisations in the experimental SMLM data-sets outnumber the modelled point-cloud by a factor of 20, this process is computationally intensive, and so these images are pre-rendered and stored on disk.

This data-set consists of 4663 individual images. Some of these show incomplete labelling or are not centriole structures (such as all the fluorophores converging on a single, bright spot). Erroneous data were removed manually, reducing the data-set size to 2055. Data was augmented by a factor of 20 rotating the entire centriole within the field of view using a 2D rotational matrix, giving a final training set size of around 40,000. As the data is represented by points and not a bitmap, it can be rotated by an arbitrary angle without introducing additional artefacts. Examples of the STORM CEP152 training images can be found in the Supplementary Data: Figure S2.

The second data-set is derived from expansion microscopy experiments to image labelled glutamylated tubulin in centrioles purified from \textit{Chlamydomoanas reinhardtii} \cite{mahecicHomogeneousMultifocalExcitation2020a}. The images are segmented and presented as tiff stacks of size 128x128x84 in xyz. A sum projection is carried out to eliminate the information in z, creating a 2D image of an unknown blur. Each image was cropped to 60x60 pixels centred on the protein complex.

As the data are represented by pixels and not a list of localisations, augmentation is limited to the four cardinal directions to avoid the creation of artefacts. The resulting data-set is 14612 items in size. As the point-spread function is not modelled, there is no base input-sigma. A Gaussian blur of decreasing sigma is applied \lq on-top-of\rq the existing image (see Supplementary Data: Figure S3).

\subsection{Input images}

The input to the network consists of a batch of 2D images, each of the same target object from different viewpoints. These images may be simulated (rendered from a known ground-truth 3D model) or derived from experimental data.

For both simulated and real SMLM data, rendering with a Gaussian generates a 2D image, with the resolution of the reconstruction being determined by the input-sigma.  For the simulated data the 2D point cloud is generated by applying a random rotation and translation, adding noise and projecting away Z.
For data in the form of images were blurred with a Gaussian, with input-sigma as the width.

Before being passed into the network, the input images were normalised to ensure that the pixel values fall within boundaries usable by the network (see Section \ref{sec:normal}).

Deep learning requires a large, representative training set for results to be accurate. For accurate 3D reconstruction, it is important to sample diverse angles since areas of the object not represented in the training data will not be reconstructed. In the simulated case, data-sets of any size can be generated (time permitting). However, this is not the case for the experimental data.

\subsection{Sigma}\label{sec:sigma}

The input-sigma value, which defines the level of blur (i.e.\ resolution) in the input images, is initialised at a high value (one which would produce an image with around diffraction limited resolution). The value is then decreased on a curve as the network trains. By starting with a larger input-sigma, the loss between the input and output images is smaller, with shallower gradients over larger distances. This allows the network to broadly optimise the points in the 3D reconstructed model matrix, refining finer detail as the input-sigma is reduced.

The lowest value for sigma can be set to the expected localization error for a particular SMLM experiment. The input-sigma curve can be found in the Supplementary Data: Figure S1. The output-sigma (that is, the sigma used by the differentiable renderer to create images from the hypothesised model) is predicted by the network. The output-sigma can be set to match the known input-sigma, but early experiments suggest that predicting the output-sigma increases the network's tolerance to scattered or missing fluorophores. By increasing the output-sigma the blur increases, accommodating the scattered points.

In experimental data, we would expect around an $\sim$8nm scatter in position due to the antibody used and an additional $\sim$12nm degradation in precision due to the localisation accuracy. Such values suggest an expected resolution around 20nm, with an expected sigma around 10nm. For our STORM CEP152 experiments we set the lower-bound of the input-sigma to a value of $\sim$3.2 pixels, which equates to 30nm using the scale provided with the data. The input-sigma changes at the end of each epoch, rather than continuously, giving a \lq stepped-curve\rq\ (see Supplementary Data: Note 3 - input-sigma Hyper-parameter and Figure S1) This is due to the images being pre-rendered before training begins. This decision was made for performance reasons.

The expansion microscopy centriole data-set has a scale of 14nm per pixel. The additional input sigma curve begins at 2.8 pixels ($\sim40$ nm), reducing to zero. The smaller initial input sigma attempts to account for the smaller image size and the unknown resolution of the data.

\subsection{Loss function}\label{sec_loss}
The loss is calculated directly between images by comparing the pixel values between the input data and the predicted result, using the PyTorch L1 Reduction option\footnote{\url{https://pytorch.org/docs/stable/generated/torch.nn.L1Loss.html}} with the \lq sum\rq \ reduction. Rather than use the L1 loss between entire images, a mask was generated from target image. The loss was calculated only for these pixels that are within the mask, with areas outside the mask set to zero for both input and output images.

\subsection{Reconstructed 3D Model Matrix and Normalisation}\label{sec:normal}
As training progresses, the matrix of 3D points that represents the reconstructed 3D model moves from a random positions to yield a recognisable structure. The size of this matrix (the number of points to optimise) is ultimately limited by the amount of memory and time available to the end user. The matrix size is chosen by the user before training starts. In simulated tests, the number of points responsible for generating the input image is generally known, except when multiple fluorophore reappearances per point are simulated.

\begin{figure}[ht!]
	\begin{center}
		\includegraphics[width=15cm]{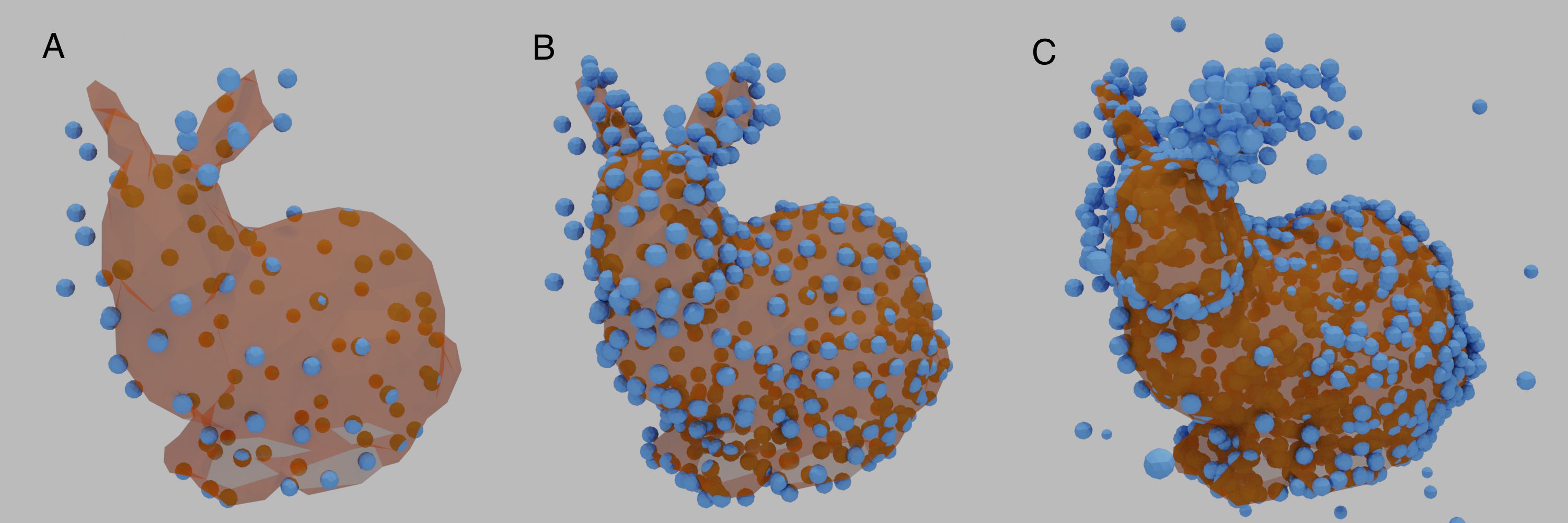}\\
	\end{center}

	\caption{Examples of a reconstructed Stanford Bunny using different sizes of 3D reconstruction matrices. Figure A contains 100 vertices. Figure B contains 350 vertices, the same number as the underlying ground truth. Figure C contains 1000 vertices. HOLLy manages to reproduce the basic shape throughout.}
\label{fig:changing_n}
\end{figure}

Since the number of points affects the integrated intensity of the image, and thus the loss, the number of points is linked to the appropriate learning rate. 
Normalisation was therefore required to bring the training data into numerical ranges the network can process without generating extreme gradients. The image tensor was divided by the integrated intensity, followed with a multiplication by a fixed scalar. Figure~\ref{fig:changing_n} shows three examples of Stanford Bunny Reconstructions, each with a different size of 3D reconstruction matrix, with normalisation applied. In each case, the basic shape is recognisable, with increasing detail.

\subsection{Hyper-parameter choices}

Hyper-parameters are the user-chosen settings \cite{goodfellowDeepLearning2016}, rather than the learned parameters. Our parameters were chosen using a combination of existing defaults and explorations within reasonable ranges.

To verify that the learning rate selected was appropriate the suggested value of 0.004 for the Adam Optimizer \cite{kingmaAdamMethodStochastic2017} was varied by a factor of 10 in both directions, stopping when structure reproduction began to fail, with a score of 0.0004. %The learning rate in the model is very closely related to the input-sigma and output-sigma values, with decreasing output-sigma resulting in an increasing error rate. 

The simulated data-sets used comprised 40,000 images, generated from an initial set of 2000 images. Each image was augmented 20 times by a random rotation around the Z axis to better match the experimental data.

The number of images presented to the network at each training step (the batch-size) can affect the final accuracy of the network \cite{kandelEffectBatchSize2020}. A batch-size of 32 was selected as appropriate. Decreasing the batch size too far caused reproduction to suffer and increasing too far caused memory usage to become computationally limiting.

The final parameter considered was the number of epochs (that is, the training time). An epoch is completed when the network has processed the entire training set once. A range of number of epochs were tested, with a value of 40 being found to be an acceptable trade-off between accuracy and time.

This baseline for training with simulated data was chosen after a number of results from earlier tests, with the restrictions of the final experimental data in mind. The most important of these is the training set size and construction. Experiments with increasing the size of the simulated training set gave improved results, but we are restricted in the size of the real, experimental data. Therefore we chose to match the size of the experimental data-set when performing the simulated experiments.

Further details of these hyper-parameters used in our experiments are listed in Supplementary Data: Note 5.

\subsection{Implementation}
Experiments were carried out with a nVidia GeForce 2080Ti GPU. Training duration was around 8 hours with the settings given as the baseline. Larger numbers of points in the reconstructed 3D model dramatically increased memory usage. 

 The estimated energy use to train a model is 2.1kWh based on a measurement of 623.4kWh over 166 days. In this period, 298 models were trained and evaluated. This was confirmed by cross-checking against the wattage of the GPU and the time spent to generate a model. 

Further technical details may be found in Supplementary Data: HOLLy Technical Details.

\section{Results}
\label{sec:results}

\subsection{Evaluation criteria}
The 3D structure which the network attempts to reconstruct is represented as a point cloud with the coordinates of each point stored in the 3D reconstruction matrix.  The network attempts to learn the orientation over time, and simultaneously improves its own internal representation of the 3D structure by comparing 2D renders of the point cloud against the training images.

The effectiveness of our approach was assessed by measuring the similarity between the input point-cloud and the resulting point-cloud stored in the model's 3D reconstruction matrix. Finding the absolute best match between two structures is an NP-hard problem, and therefore a definitive score is not possible. Given this, we selected the root mean squared distance (RMSD) between two equivalent vertices in each point-cloud as an acceptable measure. Equivalence is determined by finding the closest neighbour with the Iterative Closest Point (ICP)~\cite{arunLeastSquaresFittingTwo1987} algorithm within  CloudCompare\footnote{\url{http://www.cloudcompare.org/}}.

ICP relies on an rough, initial alignment. We performed this step manually, then applied ICP to obtain our RMSD score, independent of the pose predicted by the network. To find an RMSD score baseline to compare against we attempted to match two random clouds covering the same world-space as our model.

The parameters used in these experiments can be found in Supplementary Data: Experiment parameters.

\subsection{Simulated Results}

To assess the accuracy of our proposed method, a set of commonly used 3D models were chosen to evaluate the approach. The availability of a ground-truth structure allowed us to measure how well our network performs under different conditions. To validate our approach, we first performed a set of baseline experiments to determine how well the network could infer the 3D structure when only presented with 2D renders of these models.

\subsubsection{Baseline Experiments - Stanford Bunny}

The first model tested was the Stanford Bunny. This model has no symmetry, contains fine detail, protrusions and a homogeneous distribution of vertices across its surface. It contains considerably more points than the other point-clouds used, though the version in our experiments is in the order of hundreds of vertices as opposed to tens of thousands in the original point-cloud.

\begin{figure}[ht!]
	\begin{center}
		\includegraphics[width=15cm]{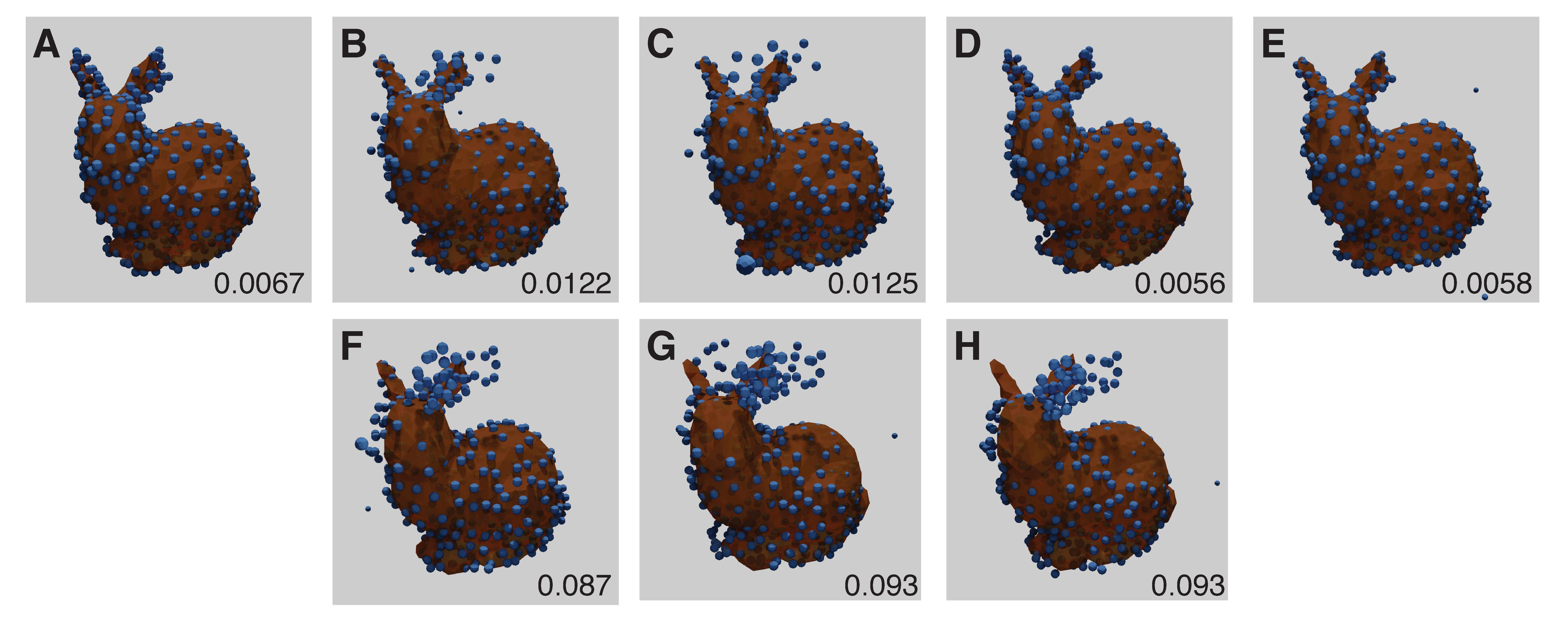}\\
	\end{center}

	\caption{Baseline Stanford Bunny experiment results for each run (A--E) with RMSD score in the lower right corner of each panel. The ground truth model is shown in orange, with the inferred structure shown as blue spheres, overlaid and aligned. (B--D) have the models mirrored for display and RMSD computation, and show high quality fitting. (F--H) correspond to results for (B--D) as the original reconstructions (without mirroring). Note that under this imaging modality, the presence or absence of mirroring cannot be determined. The parameters for this experiment can be found in Supplementary Data: Table 5.1.}
\label{fig:bl_all}\label{fig:bl_mirror}
\end{figure}

All results from baseline experiments were noise free (i.e. every generated fluorophore was exactly at an existing vertex position, there was only one per vertex position, and every vertex position was occupied). 
The baseline results all had low RMSD scores, considerably less than 0.17, the average score when aligning two random point clouds of the same size (Figure~\ref{fig:bl_all}). However, three of the runs showed a mirroring error, where the network mirrors the point-cloud in the dorsal plane. This is due to the lack of depth information in the training images (Figure~\ref{fig:bl_mirror}), and is a fundamental ambiguity.

\subsubsection{Baseline Experiments - Utah Teapot}

\begin{figure}[ht!]
	\begin{center}
		\includegraphics[width=15cm]{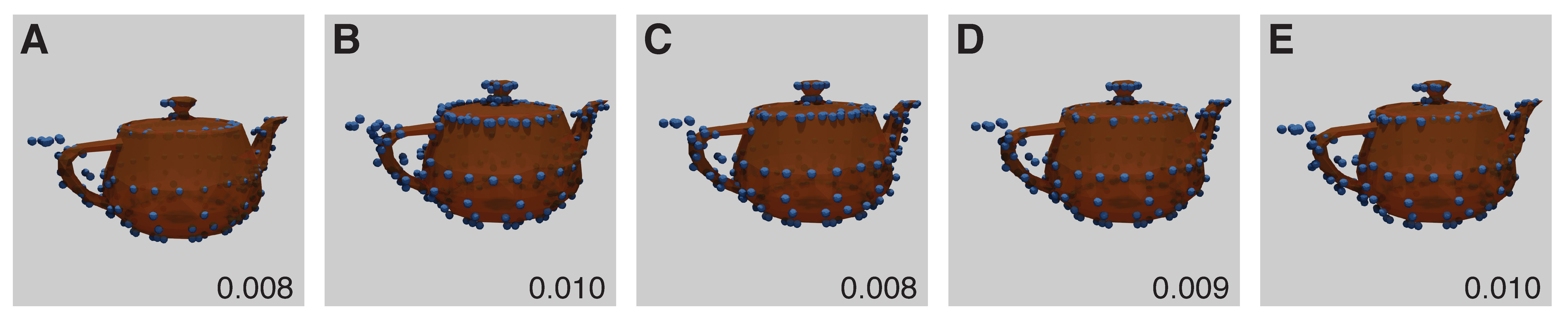}
	\end{center}
	\caption{Baseline Utah Teapot Experiment results for each run (A to E) with RMSD score. The inferred structure shown as blue spheres, overlaid and aligned against the ground truth model shown in orange. Each model shows incorrect symmetry with non-differentiated spout and handle. The parameters for this experiment can be found in Supplementary Data: Table 5.2.}\label{fig:tp_all}
\end{figure}

Our second choice of model was the Utah Teapot, which posed several challenges for our method: the similarity of the handle and spout (when rendered using points), the bilateral symmetry and the large voids between the layers of points in the central body. 

It was reconstructed well and the pose was well predicted. However, the handle appeared to be the same as the spout. Both of these areas are low in information with few ground truth points. The predicted structure therefore has an additional transverse plane of symmetry not present in the ground-truth (Figure~\ref{fig:tp_all}). From the tip of the spout, to the edge of the handle, the distance is 1036nm, using the CEP152 experiment scale.

\subsubsection{Baseline Experiments - approximation of the CEP152/HsSAS-6 complex.}

\begin{figure}[ht!]
	\begin{center}
		\includegraphics[width=15cm]{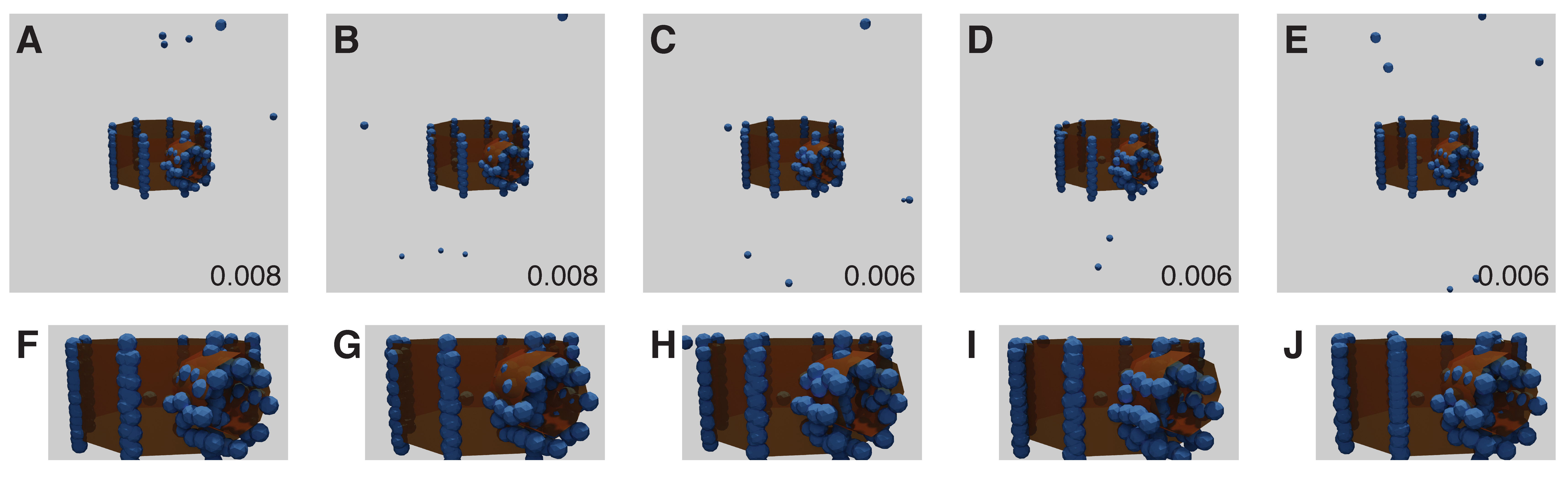}
	\end{center}
	\caption{Baseline CEP152/HsSAS-6 approximation experiment results for each run (A--E, top row) with RMSD score. The ground truth model is shown in orange, with the inferred structure shown as blue spheres, overlaid and aligned.\label{fig:sus_all}
    Bottom row (F--J): close up of the top row. Note the slight offset of the smaller, reconstructed cylinder from the ground-truth. The parameters for this experiment can be found in Supplementary Data: Table 5.3.
	\label{fig:sus_close_all}}
\end{figure}

The third point-cloud used in these experiments is an approximation of the CEP152/HsSAS-6 complex \cite{siebenMulticolorSingleparticleReconstruction2018}. The approximation consisted of two cylinders, one smaller and perpendicular to the other. This point-cloud is somewhat smaller than the others and is extremely regular with large gaps between the columns of points.

The smaller, cylindrical structure is offset towards the top of the larger structure in the ground-truth; this is not reflected in the reconstruction. This is likely due to the size of the point-cloud in the view - fine detail is hard to discern when the point-cloud is small (Figure~\ref{fig:sus_close_all}). From the end of the small cylinder to the furthest edge of the larger cylinder, the distance is roughly 415nm.

Together, these baseline experiments indicate that our approach is suitable for reconstructing the overall 3D structure from a series of 2D images. Most results showed low RMSD scores and produced structures that are a good match to the original 3D models.

\subsection{Modelling Experimental Noise in Simulated Results}

Our method aims to discern structure from fluorescence microscopy images, particularly super-resolution. We therefore focused on the kinds of problems often encountered in such experiments. Fluorophores are offset from the object they are labelling, they may not bind to certain areas, or might bind multiple times. They may not illuminate consistently or they may not be separable from their neighbours. We modelled three forms of experimental noise: missing fluorophores (where no fluorophores appear for a particular ground-truth point), scatter (where a fluorophore appears at a varying distance from its ground-truth point), and multiple binding (where multiple fluorophores appear for a single ground-truth point).

\subsubsection{Scatter}

\begin{figure}[ht!]
	\begin{center}
			\includegraphics[width=15cm]{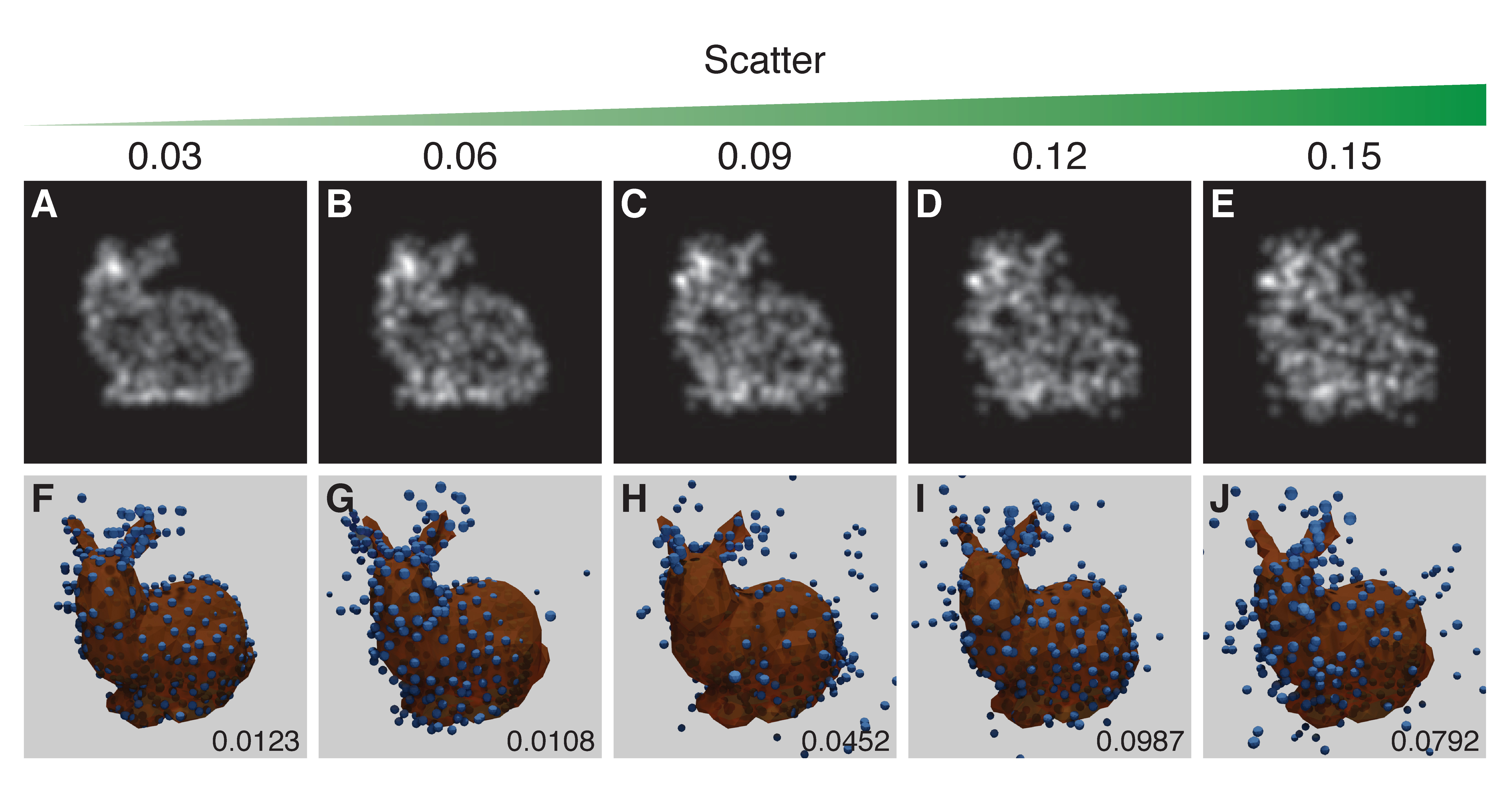}
	\end{center}
	\caption{The results of the experiment into the effect of scatter. Top row (A--E): examples of training images treated with increasing scatter, as indicated by the scatter value above each panel. Bottom row (F--J): the corresponding results with the inferred structure shown as blue spheres, overlaid and aligned with the ground truth model shown in orange. RSMD scores are indicated for each run in the lower right corner of (F--J). Runs in (F) and (C) respectively showed an incorrect symmetry in structure and mirroring in a vertical plane. The parameters for this experiment can be found in Supplementary Data: Table 5.4.}\label{fig:scatter_all}
\end{figure}

Two factors can lead to scatter in fluorophore positions: the inaccuracy of the fitted position due to the limited number of photons collected, and the offset between the protein of interest and the label, with the largest effect arising from primary/secondary antibody labelling.
This noisy change of position (scatter) is modelled using a random Gaussian distribution with a particular scatter-sigma value. The scatter-sigma ranges from 0.03 pixels to 0.15 pixels (9~nm to 44~nm, given the scale in the CEP152 experimental data).

The results suggest that a scatter-sigma value between 0.06 pixels and 0.09 pixels (20~nm to 29~nm) is the cut-off point for acceptable reproduction of the structure. The run in Figure~\ref{fig:scatter_all}F shows a rare error where the structure is symmetrical along the dorsal plane - effectively giving the structure two heads. Figure~\ref{fig:scatter_all}C suffers from the mirroring problem (Figure~\ref{fig:scatter_all}).

\subsubsection{Missing fluorophores}

\begin{figure}[ht!]
	\begin{center}
		\includegraphics[width=15cm]{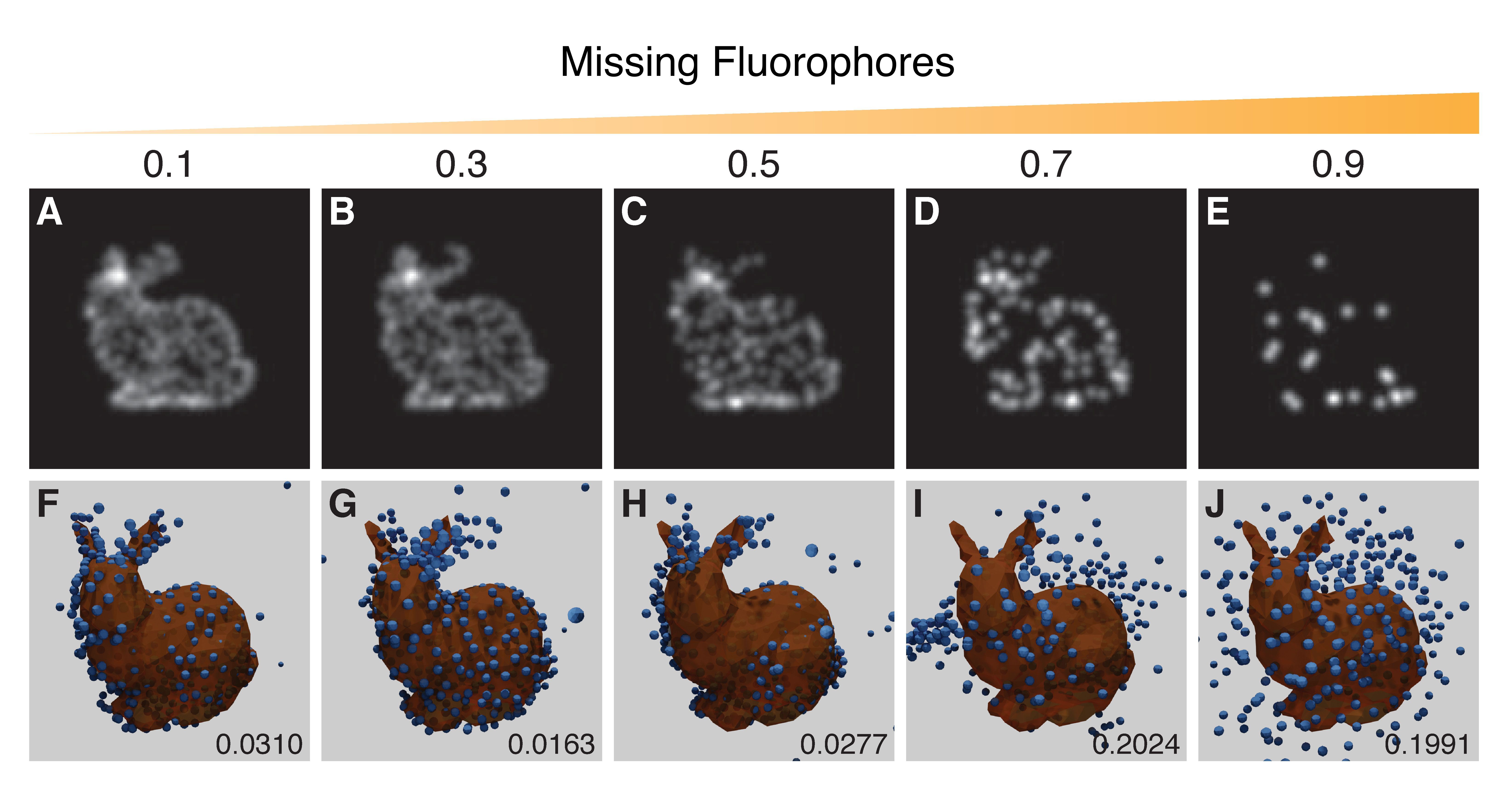}
	\end{center}
	\caption{Results from the experiment on the effect of missing fluorophores. Top row (A--E): examples of training images with increasing probability of removing points as indicated by the value above each panel. Bottom row (F--J): the corresponding results with the inferred structure shown as blue spheres, overlaid and aligned with the ground truth model in orange. Resulting RSMD scores for each run are indicated in the lower right corner of (F--J). The run in (G) showed an incorrect symmetrical structure whereas the run in (H) showed mirroring in a vertical plane. The parameters for this experiment can be found in Supplementary Data: Table 5.5.}\label{fig:dropout_all}
\end{figure}

When a fluorescence microscopy sample is labelled, not every potential site is labelled, and not all fluorophores will fluoresce. The degree of labelling and the performance of fluorophores strongly impacts image quality. To simulate this effect
a random selection of vertex positions are not labelled with fluorophores.
Results suggest that a recognizable reproduction with a good RMSD score can be obtained with up to $\sim$30\% of the points removed (Figure~\ref{fig:dropout_all}).

\subsubsection{Multiple binding and Scatter}

\begin{figure}[ht!]
	\begin{center}
		\includegraphics[width=15cm]{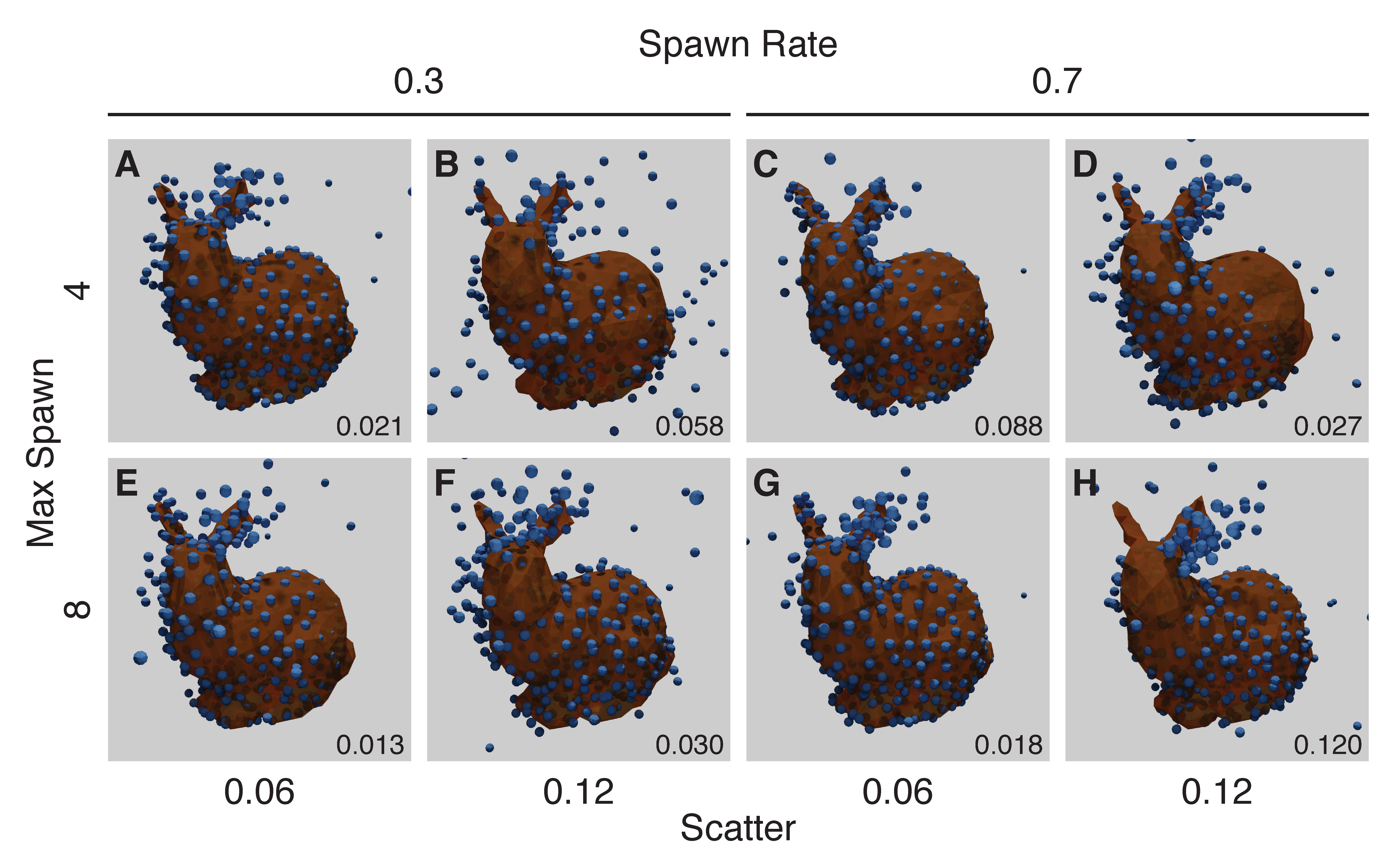}
	\end{center}
	\caption{Results from the Noise Experiment for each run (A to H), with resulting RMSD shown in the lower right corner of each panel. The ground truth model is shown in orange, with the created structure shown as blue spheres, overlaid and aligned. The top row (A--D) shows runs with a maximum number of flurophores per ground-truth point of 4, while the bottom row (E--H) shows runs with a maximum of 8. The left two columns (A, B, E, F) have a spawn-rate of 0.3, with the right two columns (C, D, G, H) have a spawn-rate of 0.7. Runs in (A, C, D, E, F, G) have incorrect symmetry whereas the run in (H) has mirroring in a vertical plane. The parameters for this experiment can be found in Supplementary Data: Table 5.6.}\label{fig:noise_all}
\end{figure}
Our final noise experiment randomly chooses up to a maximum number of bound fluorophores per ground-truth point, each with a random scatter. A single ground-truth point may \lq spawn\rq \ up-to a maximum of individual fluorophores (\emph{max-spawn}) using a user-set probability (\emph{spawn-rate}). In these experiments we chose a number of parameters for \lq max-spawn\rq, \lq spawn-rate\rq and scatter.

Many of these runs show symmetrical structure where none should occur (Figure~\ref{fig:noise_all}), in a manner similar to the missing fluorophores experiment (Figure~\ref{fig:dropout_all}).

These results provide additional confidence that HOLLy can produce accurate structures from experimental data. The majority of results have low RMSD scores, with identifiable structures and some tolerance to noise.

\subsection{SMLM dataset of the CEP152 Complex}
Having optimized our approach with different 3D models, we next applied it to experimental SMLM data collected on the CEP152 complex, which is part of the centriole. One important factor with this data is that the integrated intensity varies considerably across the CEP152 data-set, with the number of localisations ranging from 5,000 to 30,000. Normalisation plays a key part in making sure this intensity range can be modelled by our network. Additionally, since this data-set is limited by the number of feasible experiments, data-augmentation plays a key role in increasing both the absolute number of training images and the variety of orientations. This training data-set consists of approximately 40,000 images, augmented from an experimental data-set of approximately 2000 images. See Supplementary Data: Figure S2 for representative images that illustrate the range of orientations and experimental noise in this training data-set.

\begin{figure}[ht!]
	\begin{center}
		\includegraphics[width=15cm]{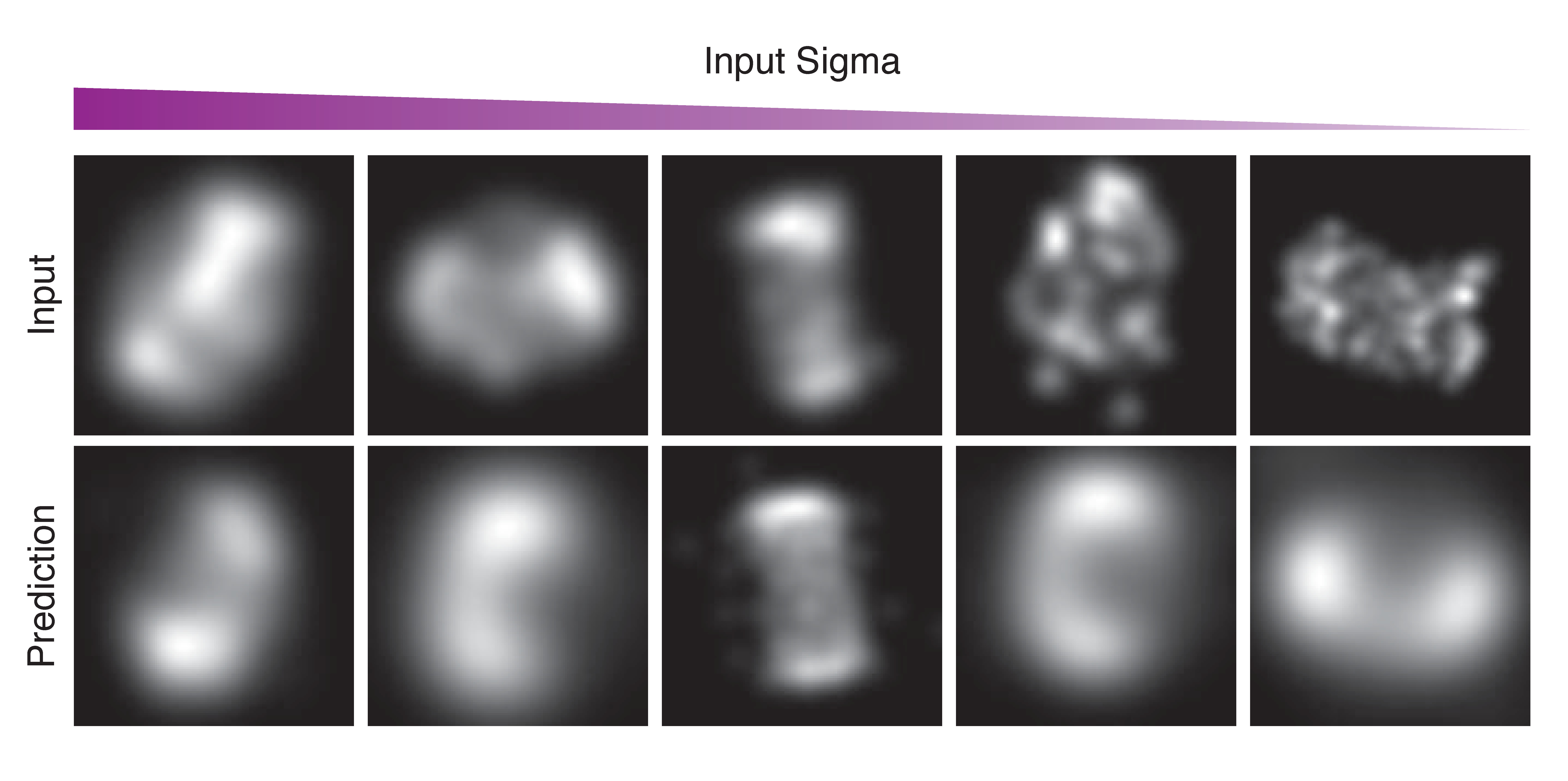}
	\end{center}
	\caption{Examples from the first run of the STORM CEP152 data-set, rendered at different points during training as input-sigma values decreased. The top row shows input images from the test set. The bottom row shows the corresponding prediction. }\label{fig:cep_pred}
\end{figure}

\begin{figure}[ht!]
	\begin{center}
		\includegraphics[width=15cm]{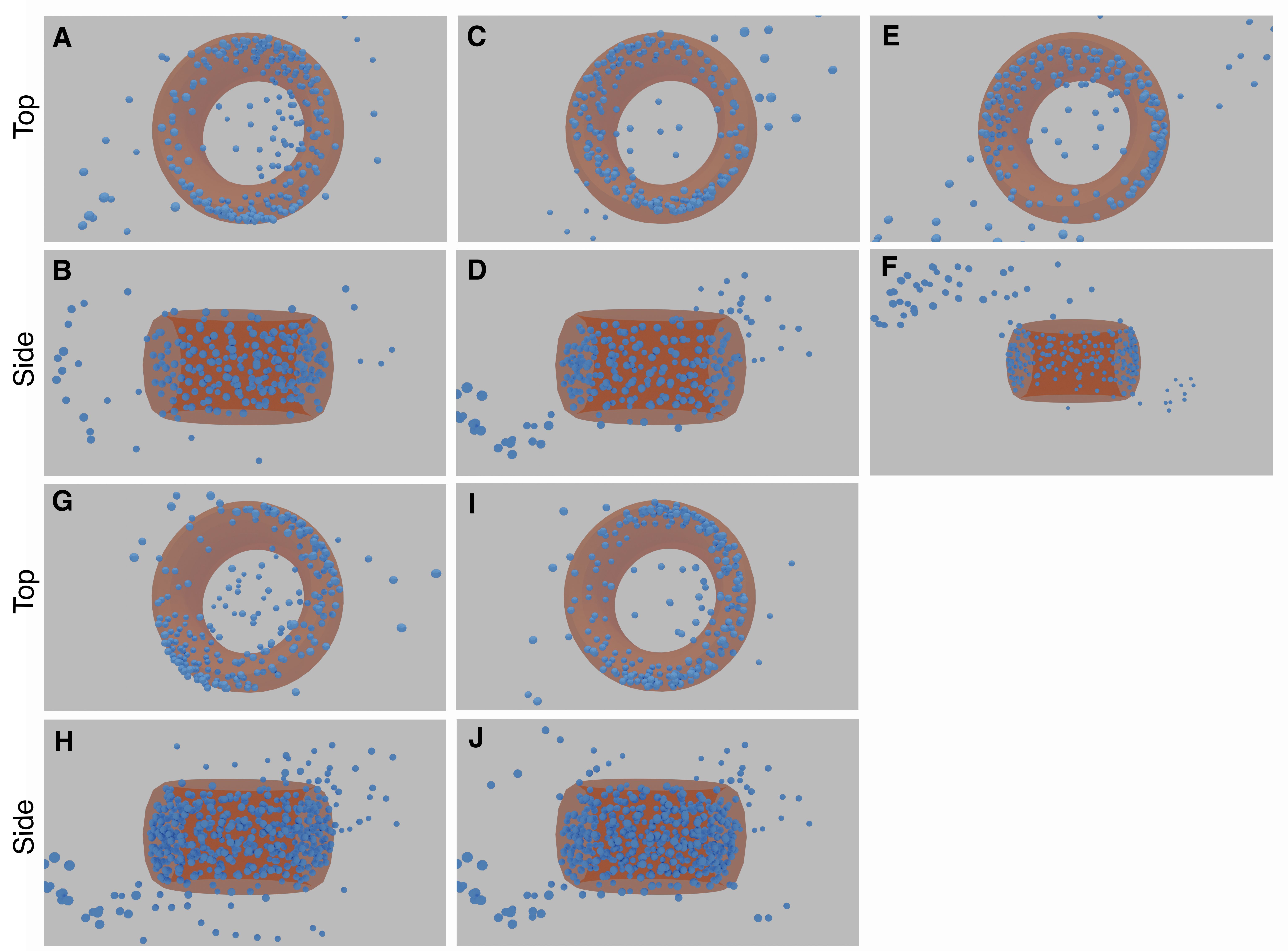}
	\end{center}
	\caption{Results of the five STORM CEP152 experiments (each with a top and side view image pair). The first experiment comprises image A and B, the second experiment C and D, and so forth. The torus structure within the blue point cloud is highlighted with an orange ring. See Supplementary Data - Video 1 and Video 2 for greater clarity.}\label{fig:cep_all}
\end{figure}

After training with these images, our network converged on a central torus for the CEP152 complex (Figures~\ref{fig:cep_pred}-\ref{fig:cep_all}). This inferred structure is consistent with the confirmed structure of this protein complex \cite{siebenMulticolorSingleparticleReconstruction2018, kimMolecularArchitectureCylindrical2019}.

Figure \ref{fig:cep_pred} in particular, shows examples of the network attempting to match the training images, both in terms of structure and the input-sigma. The input images are not completely static; recall they are generated with a particular input-sigma, which decreases as training progresses. However, the output-sigma predicted by the network does not continuously decrease as the input-sigma does - rather the rate begins to flatten towards the end of training. Indeed, certain images are rendered with a higher blur than others, suggesting that certain images are being compensated for with a higher output-sigma.

The final 3D structures in Figure~\ref{fig:cep_all} can be seen more easily in the videos which accompany this paper (see Supplementary Data: Video 1 and Video 2). When rendering these predicted structures in 2D based on the inferred orientations, they show significant blurring due to a large predicted output-sigma, even when the input-sigma was low (Figure~\ref{fig:cep_pred}). There was some noise in the inferred structure, with two \lq fringe-like\rq \ structures in some of the runs (Figure~\ref{fig:cep_all}).  Some points still appear in the middle of the toroidal structure, likely because the network has been unable to optimise these points as any direction they might now move in would result in an increasing error. The density appears to be lower for a small arc on the torus, reflective of the input images that also show a similar effect. These final structures are not exact as some noise still remains. Nonetheless, the consensus result that emerges from multiple runs is a toroidal structure that matches that of the CEP152 complex.

\subsection{SIM/expansion microscopy dataset of  glutamylated tubulin in centrioles}

\begin{figure}[ht!]
	\begin{center}
		\includegraphics[width=15cm]{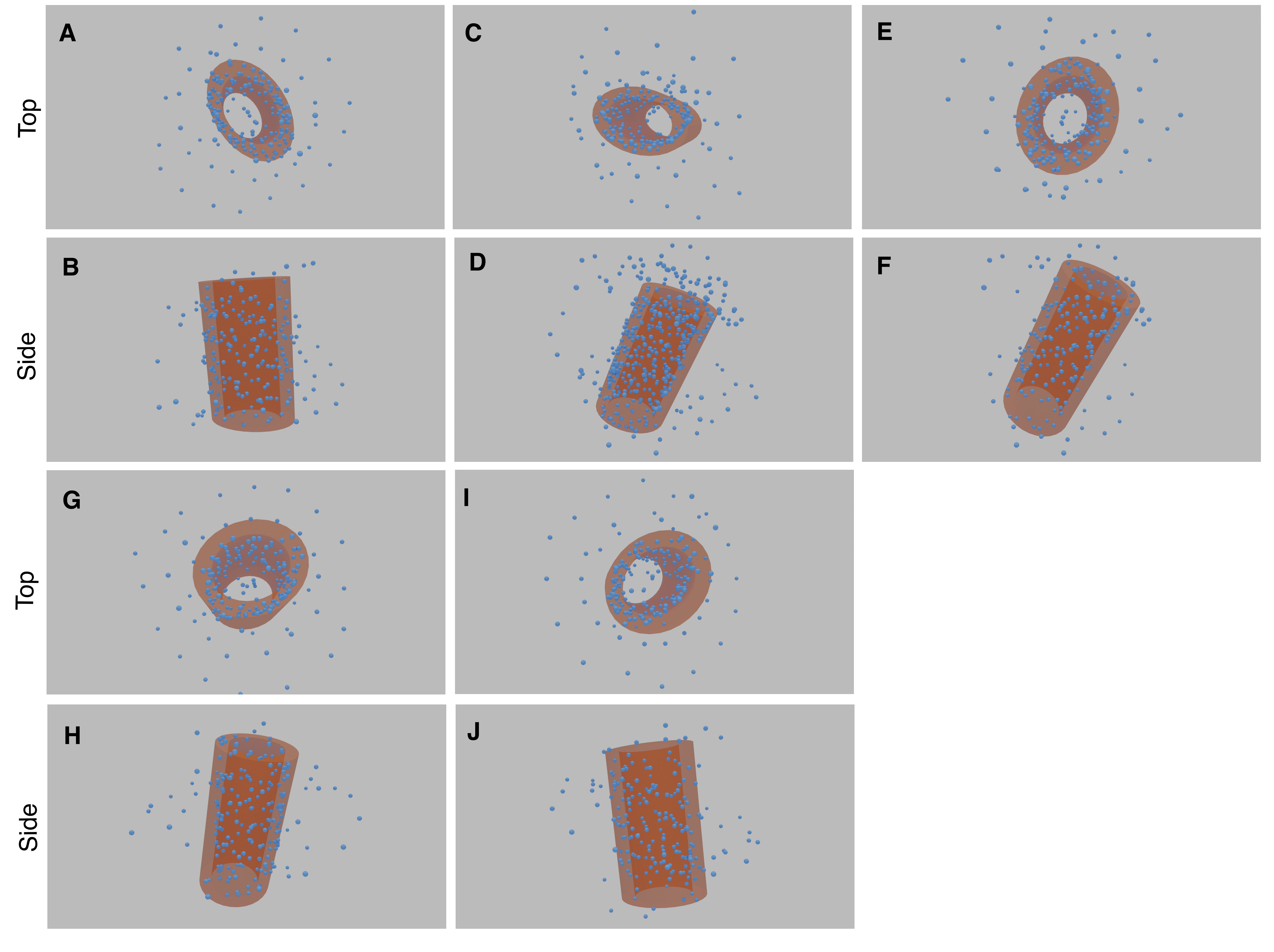}
	\end{center}
	\caption{Results of the five experiments on SIM/expansion Microscopy of glutamylated tubulin in purified centrioles (each with a top and side view image pair). The first experiment comprises image A and B, the second experiment C and D, and so forth. The cylinder structure within the blue point cloud is highlighted with an orange cylinder. See Supplementary Data - Video 3 and Video 4 for greater clarity.}\label{fig:dora_all}
\end{figure}

To validate our method, we also applied it to a separate experimental data-set~\cite{mahecicHomogeneousMultifocalExcitation2020a} obtained using a different imaging technique. We analysed SIM / expansion microscopy images of glutamylated tubulin in purified centrioles. After training on the SIM/expansion microscopy images, our network converged on a central cylinder for this complex (Figure~\ref{fig:dora_all}). The density of points is highest in the centre of each image, with a tube-like structure visible. These 3D aspects are clearer in the supplementary Data: Video 3 and Video 4. There appears to be a  \lq frill-like\rq structure around the top to middle of the cylinder, which may reflect a particular characteristic of the input data. Many of the images show a spike like protrusion, emanating from the top of the central cylinder (see Supplementary Data: Figure S3).

The consensus elongated cylindrical structure produced by our method is also consistent with the known structure of glutamylated tubulin in centrioles \cite{mahecicHomogeneousMultifocalExcitation2020a} (See Supplementary Data: Figure S2).

\subsection{Handedness}

Often when reconstructing 3D shapes from macroscopic images, perspective
projection and occlusion effects can be used to infer depth. Neither of these are present in 2D fluorescence microscopy images. Without perspective projection, there is an unknown reflection of the final 3D geometry which cannot be determined from the data. This is known as the affine ambiguity~\cite{hartleyMultipleViewGeometry2004}.
Examples of this effect can bee seen in Figure~\ref{fig:bl_mirror}.

\section{Discussion}

We have demonstrated a method that enables 3D structures to be reconstructed from sets of 2D SMLM or fluorescence microscopy images without any template or symmetry constraints.
Our method, HOLLy, can tolerate both scatter and the limited labelling efficiency of experimental fluorescence images. The training process results in a 3D model of the structure encoded as a point-cloud in the 3D reconstruction matrix. Based on estimates of RMSD values against ground-truth and visual inspection of the results, we find that our approach can create accurate reconstructions of 3D macro-molecular structures.

Our results also demonstrate the limitations of the technique. Because of the use of 2D images, the technique is unable to resolve the chirality of the model. In addition, when the data quality is poor small structures are not reproduced.
As a result when the structure is close to symmetric, the final model may become actually symmetric. On experimental data, the presence of these issues could potentially be identified by training on the same data-set multiple times and examining the differences between the results.

The value of reconstructing multiple images of a structure into a single hypothesised structure has been demonstrated in cryo-EM. In SMLM such approaches exist \cite{heydarianThreeDimensionalParticle2019}, and show an improvement in the signal to noise ratio when combining multiple images, but performing such fits on complex structures with no constraints is extremely challenging. Here we show that, by building a 3D model and using a neural network for predicting rotation, HOLLy can discern structure from localisations with a data-set of 2000 unique images. With the increased popularity of high throughput SMLM techniques~\cite{holdenHighThroughput3D2014,barentine3DMulticolorNanoscopy2019}, HOLLy provides a way to extract structural information from large volumes of super-resolution microscopy data without assumptions.

\section*{Conflict of Interest Statement}
Author Ed Rosten is employed by Snap Incorporated. The remaining authors declare that the research was conducted in the absence of any commercial or financial relationships that could be construed as a potential conflict of interest

\section*{Author Contributions}

Initial conception and design: SC, ER. Acquired high-throughput SMLM images: CS, SM. Performed data cleaning, deep learning implementation, experiments: BB. Supervised the research: SC, ER, QC.  Interpreted and discussed results: BB, SC, ER, QC, CS, SM. Comments on the manuscript: CS, SM. Wrote the paper: BB, SC, ER, QC. 

\section*{Funding}

BB is funded by a studentship from the UKRI/BBSRC National Productivity Investment Fund (BB/S507519/1) and is part of the London Interdisciplinary Doctoral Programme funded by UKRI/BBSRC (BB/M009513/1).

\section*{Data Availability Statement}
\label{sec:data}
The program HOLLy can be found at \url{https://github.com/OniDaito/Holly.git}. The simulated data results presented in the paper can be reproduced with the configuration files found in that repository.

The STORM CEP152 complex data used in our experiments can be found on Zenodo at \url{https://doi.org/10.5281/zenodo.4751056}. Once downloaded and extracted, our results can be reproduced using the configuration file found in the HOLLy project.

The GT335 tublin, centriole data can be found at \url{https://10.5281/zenodo.3613906}.

The results from out CEP152 experiments we have presented can be downloaded from Zenodo at \url{https://zenodo.org/record/4836173}.

\bibliographystyle{plain}
\bibliography{paper_export}
	
\end{document}

\onecolumn

\maketitle

\section{HOLLy Technical Details}

Our method relies upon three major components: a convolutional neural network, a vertex tensor representing the structure used internally and a differentiable renderer. The neural network is responsible for generating the parameters required by the renderer from a batch of input images. These parameters are combined with the set of points held in the vertex tensor and passed to the renderer, whereupon an output image is produced. The differentiable renderer is a simplified 3D rasterisation pipeline of transformation matrices, though the perspective matrix is omitted as our microscopy images do not show any apparent change due to distance from the camera. Several advanced, differentiable renderers already exist though they are designed to render polygons instead of Gaussian-blurred points. This set of transformation matrices is completely differentiable allowing for error gradients to pass backward from the loss function through the pipeline.

Neural networks learn through back-propagating the error (or loss) between the expected output and the generated output. This loss is produced by a loss function - typically the Least Squared Error (or L2) loss.  We chose the Least Absolute Deviations (or L1) loss on the pixel values of the input and output image batches. L1 was chosen as it has been shown to outperform the more common L2 on pixel comparison tasks. The loss is sent to the Adaptive Moment Estimation (Adam) stochastic optimiser \cite{kingmaAdamMethodStochastic2017} in order to update the network weights and the positions of the points in our structure set.

Our network consists of 10 convolutional layers and two fully connected layers. Leaky ReLU (Rectified Linear Unit) is the activation function used at each layer except the final output layer. The output layer comprises the rotational and translational parameters, and in certain experiments, the predicted sigma value, discussed in the experiment section. Each convolutional layer uses batch normalisation \cite{ioffeBatchNormalizationAccelerating2015a}. The first convolutional kernel is 5 by 5 pixels, with all subsequent kernels being 3 by 3 pixels. A padding size of 2 and a stride of 2 are used. After the first layer, the subsequent layers come in pairs - the first having a stride of two, the second having a stride of 1. This performs a similar operation to a pooling layer. The final convolutional layer performs a final pooling with it's stride of 2.

The vertex tensor consists of a number of points in 3D space, represented in homogeneous coordinates (x, y, z and w). The size of this set is chosen by the user to best match the integrated density of the input data. When no normalisation strategy is applied, each individual point has an intensity of 1. Network performance is highly dependent on the size of this set. Therefore there is a trade-off between model accuracy and speed, with memory requirements scaling significantly with the number of points requested. The number of points is a so-called \lq hyper-parameter\rq, along with other hyper-parameters such as the learning rate and batch size.

The network predicts between 3 and 6 continuous parameters - depending on the options chosen by the user. The orientation is represented in \lq Angle-Axis\rq \ format:
3 real numbers for the rotation axis with the magnitude representing the angle. These are unchanged and passed to the renderer directly from the neural network. Translation in the X-Y plane is restricted via a \emph{tanh} function, which has the effect of limiting very large movements. Translation is restricted to -0.1 to 0.1. The final parameter is the sigma value used to render the output images. Output sigma is clamped to a maximum of 14 and is passed through a Softplus function.

The output of the renderer is a batch of images. Each fluorophore in the image is represented as a 2D Gaussian with a particular sigma (Equation \ref{eq_gaussian}). The input data is rendered with a sigma value chosen by the user. The output data generated by the network can either have its sigma value derived by the network or also set by the user. A large sigma - for example 10.0 - results in a very blurry image, with each point having a Full Width at Half Maximum (FWHM) of 138 pixels, whereas a sigma of 1.2 has a FWHM of 2 pixels. The network begins training on images rendered with a large sigma, progressing to a smaller sigma as training continues. This leads to better convergence as large changes in pixel intensity across the image are not present, providing smoother gradients over large distances. Initially, points are able to move further, refining their positions with smaller movements later in the training.

\begin{equation} \label{eq_gaussian}
G(x, y) = \frac{1}{2\pi\sigma^{2}}e^{-\frac{x^{2} + y^{2}}{2\sigma^{2}}}
\end{equation}

\section{Converting MATLAB SMLM data}

The program used to convert the experimental STORM CEP152 data into a series of FITS images can be found on GitHub\footnote{ \label{ceprender_url}\url{https://github.com/onidaito/ceprender/}}. It uses the same rendering pipeline as the differentiable renderer found in HOLLy.

Generating a set of training images to replicate our results takes roughly 3 days on a AMD Ryzen Threadripper 1920X (24) running at 3.500GHz.

\section{Input-sigma Hyper-parameter.}

\begin{figure}
	\centering
	\includegraphics[width=0.8\textwidth]{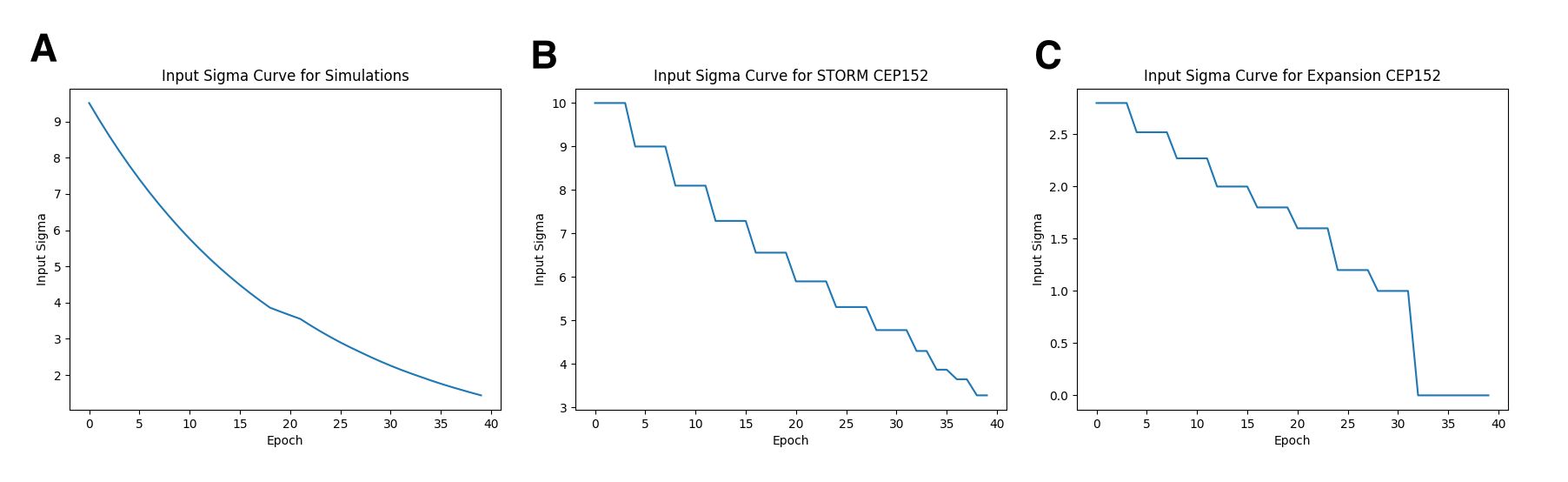}
	\caption{The input-sigma curve used in our experiments with simulated data models (A), STORM CEP152 data (B) and Expansion Microscopy CEP152 (C). The input-sigma in image C is the level of blur applied to images that already contain some unknown level of blur. The final value of 0 in image C is intentional; the images are being use with no additional blur.} 
	\label{fig:sigma_curves}
\end{figure}

The two input-sigma profiles are shown in Figure \ref{fig:sigma_curves}. We generated images on demand through training when using simulated point-clouds, therefore the input curve can be smoothly altered per training step as opposed to per-epoch. However, the experimental CEP152 data was rendered with discrete sigma values, therefore has a \lq stepped\rq \ curve.

%% More detail in the figure caption here

\section{Final structure videos.}
To aid in visualising the final 3D structure of the STORM CEP152 complex as discerned by HOLLy, we have included a number of videos that show the final structure for each of the 5 experiments (Video S1), including the progression through training (Video S2).

Similarly, to aid in visualising the final 3D structure of the Expansion Microscopy Centrioles as discerned by HOLLy, we have included a number of videos that show the final structure for each of the 5 experiments (Video S3), including the progression through training (Video S4).

\begin{figure}
	\centering
	\includegraphics[width=0.8\textwidth]{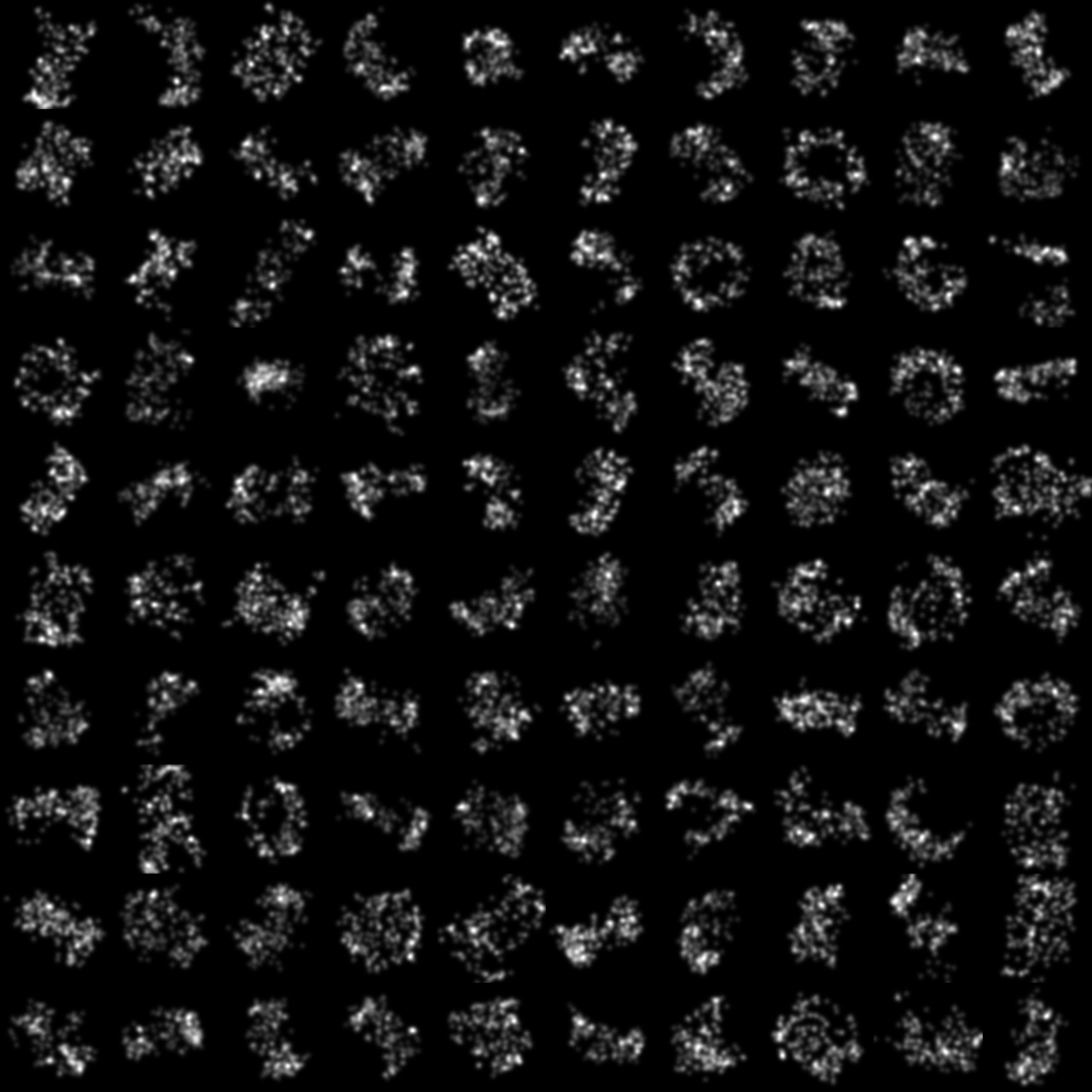}
	\caption{A random selection of 100 images from the STORM CEP152 dataset, rendered with an input sigma of 2.52. This small sample is broadly representative of the 40,000 individual images within the training dataset.} 
	\label{fig:training}
\end{figure}

\pagebreak

\begin{figure}
	\centering
	\includegraphics[width=0.8\textwidth]{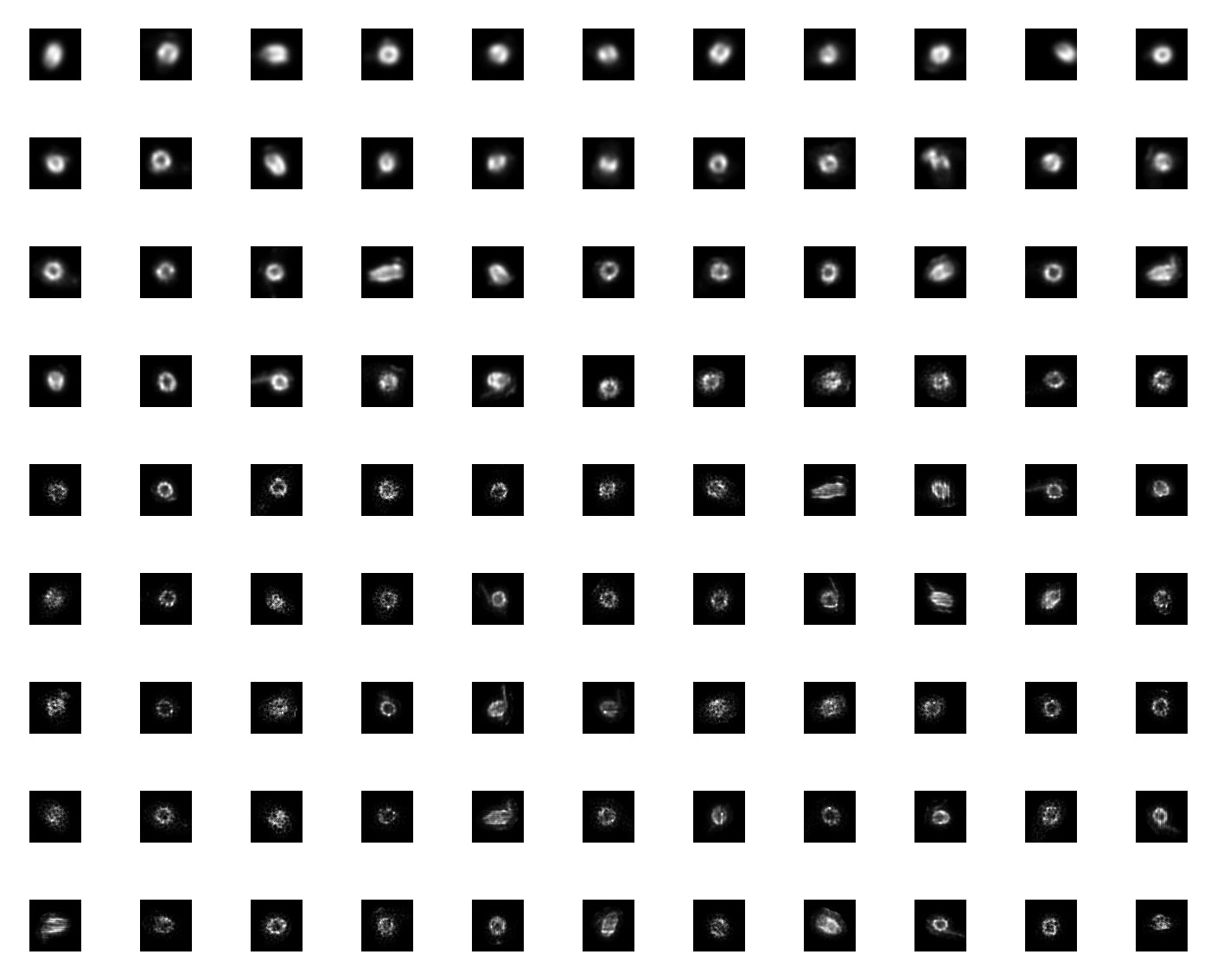}
	\caption{A random selection of 100 images from the Expansion Microscopy Centriole dataset. This small sample is broadly representative of the 14,000 individual images within the training dataset. Note that these images are much smaller than these used in the simulated or STORM experiments.} 
	\label{fig:training_dora}
\end{figure}

\pagebreak

\section{Experiment parameters}

The key parameters for every experiment are listed below. Example parameters for the baseline experiments are included in the code-base, available online (see Code Availability Statement).\\

\subsection{Stanford Bunny Experiment parameters}
\begin{center}
 \begin{tabular}{|c c c c c c|} 
 \hline
 Run & Epochs & Learning Rate & Num. of Points & Max. Input Sigma & Min Input Sigma \\ [0.5ex] 
 \hline
 1.A to 1.E & 40 & 0.0004 & 350 & 10 & 1.41 \\ 
 \hline
\end{tabular}
\end{center}

\subsection{Utah Teapot Experiment parameters}
\begin{center}
 \begin{tabular}{|c c c c c c|} 
 \hline
 Run & Epochs & Learning Rate & Num. of Points & Max. Input Sigma & Min Input Sigma \\ [0.5ex] 
 \hline
 2.A to 2.E & 40 & 0.0004 & 230 & 10 & 1.41 \\ 
 \hline
\end{tabular}
\end{center}

\subsection{CEP152 Approximation Experiment parameters}
\begin{center}
 \begin{tabular}{|c c c c c c|} 
 \hline
 Run & Epochs & Learning Rate & Num. of Points & Max. Input Sigma & Min Input Sigma \\ [0.5ex] 
 \hline
 3.A to 3.E & 40 & 0.0004 & 180 & 10 & 1.41 \\ 
 \hline
\end{tabular}
\end{center}

\subsection{Scatter Experiment parameters}
\begin{center}
 \begin{tabular}{|c c c c c c c|} 
 \hline
 Run & Epochs & Learning Rate & Num. of Points & Scatter & Max. Input Sigma & Min Input Sigma \\ [0.5ex] 
 \hline
 4.A & 40 & 0.0004 & 350 & 0.03 & 10 & 1.41 \\
 4.B & 40 & 0.0004 & 350 & 0.06 & 10 & 1.41 \\
 4.C & 40 & 0.0004 & 350 & 0.09 & 10 & 1.41 \\
 4.D & 40 & 0.0004 & 350 & 0.12 & 10 & 1.41 \\ 
 4.E & 40 & 0.0004 & 350 & 0.15 & 10 & 1.41 \\ 
\hline
\end{tabular}
\end{center}

\subsection{Drop-out Experiment parameters}
\begin{center}
 \begin{tabular}{|c c c c c c c|} 
 \hline
 Run & Epochs & Learning Rate & Num. of Points & Drop-out & Max. Input Sigma & Min Input Sigma \\ [0.5ex] 
 \hline
 5.A & 40 & 0.0004 & 350 & 0.1 & 10 & 1.41 \\
 5.B & 40 & 0.0004 & 350 & 0.3 & 10 & 1.41 \\
 5.C & 40 & 0.0004 & 350 & 0.5 & 10 & 1.41 \\
 5.D & 40 & 0.0004 & 350 & 0.7 & 10 & 1.41 \\ 
 5.E & 40 & 0.0004 & 350 & 0.9 & 10 & 1.41 \\ 
\hline
\end{tabular}
\end{center}

\subsection{Spawn Experiment parameters}

\begin{center}
 \begin{tabular}{|c c c c c c c c|} 
 \hline
 Run & Epochs & Learning Rate & Num. of Points & Drop-out & Scatter & Max Spawn & Spawn Rate \\ [0.5ex] 
 \hline
 6.A & 40 & 0.0004 & 350 & 0.1 & 0.06 & 4 & 0.3 \\
 6.B & 40 & 0.0004 & 350 & 0.1 & 0.12 & 4 & 0.3 \\
 6.C & 40 & 0.0004 & 350 & 0.1 & 0.06 & 4 & 0.6 \\
 6.D & 40 & 0.0004 & 350 & 0.1 & 0.12 & 4 & 0.6 \\ 
 6.E & 40 & 0.0004 & 350 & 0.3 & 0.06 & 4 & 0.3 \\
 6.F & 40 & 0.0004 & 350 & 0.3 & 0.12 & 4 & 0.3 \\
 6.G & 40 & 0.0004 & 350 & 0.3 & 0.06 & 4 & 0.6 \\
 6.H & 40 & 0.0004 & 350 & 0.3 & 0.12 & 4 & 0.6 \\
\hline
\end{tabular}
\end{center}

\section{Direct Optimisation}

We tested whether or not we could directly optimise the pose parameters with no convolutional layers. We used the Utah Teapot and Stanford bunny models, no sigma prediction and the following parameters:

\begin{center}
 \begin{tabular}{|c c c c c c c c|} 
 \hline
 Run & Model & Epoch & Learning Rate & Num. of Points & Max. Input S. & Min Input S. & Optimiser \\ [0.5ex] 
 \hline
 7.A & teapot & 100 & 0.0004 & 250 & 10 & 1.41 & Adam \\
 7.B & teapot & 100 & 0.004 & 250 & 10 & 1.41 & Adam \\
 7.C & bunny & 100 & 0.0004 & 350 & 10 & 1.41 & Adam \\
 7.D & bunny & 100 & 0.004 & 350 & 10 & 1.41 & Adam \\
 7.E & teapot & 100 & 0.0004 & 250 & 10 & 1.41 & SGD \\
 7.F & teapot & 100 & 0.004 & 250 & 10 & 1.41 & SGD \\
 7.G & bunny & 100 & 0.0004 & 350 & 10 & 1.41 & SGD \\
 7.H & bunny & 100 & 0.004 & 350 & 10 & 1.41 & SGD \\
 \hline
\end{tabular}
\end{center}

\pagebreak
We also performed a run with noise added, using the following parameters:

\begin{center}
 \begin{tabular}{|c c c c c c c c c|} 
 \hline
 Run & Spawn Rate & Scatter & Epoch & L. Rate & Num. of Points & Max. Input S. & Min Input S. & Optimiser \\ [0.5ex] 
 \hline
 7.I & 0.3 & 0.06 & 100 & 0.0004 & 350 & 10 & 1.41 & SGD \\

 \hline
\end{tabular}
\end{center}

\begin{figure*}[b]
	\centering
	\includegraphics[width=0.8\textwidth]{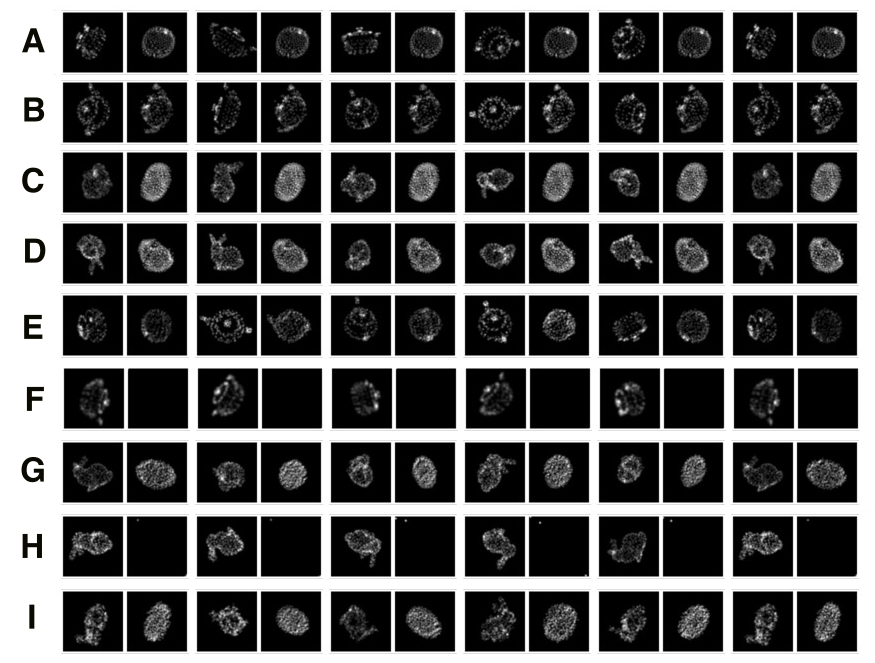}
	\caption{The final results from runs 7.A to 7.I - direct optimisations. Each row contains 6 pairs of images. The first image in each pair is the input image - the second is the corresponding output. Run 7.B appears to show a derived structure similar to the ground-truth, though this is misleading as the model only appears reasonable from one angle (viewing the final model in 3D shows this more clearly). Runs 7.F and 7.H fail to create any structure. Runs 7.E, 7.G and 7.I show some response to the different input rotations, whereas the remaining runs show very similar output poses.} 
	\label{fig:directA}
\end{figure*}

\bibliographystyle{plain}
\bibliography{paper_export}